\documentclass{article}
\usepackage{amsmath}
\usepackage{amssymb}
\usepackage{epsfig}
\usepackage{hyperref}
\usepackage{graphicx}
\usepackage{epstopdf}
\usepackage{amsmath}
\usepackage{setspace}
\usepackage{mathrsfs}
\usepackage{subfig}
\usepackage{color}
\usepackage{mathtools}
\usepackage{lipsum}
\usepackage[affil-it]{authblk}
\usepackage{cite}
\usepackage{leftidx}
\DeclarePairedDelimiter{\ceil}{\lceil}{\rceil}
\usepackage[pagewise]{lineno}

\begin{document}

\title{Analysis of dispersion and propagation properties in a periodic rod using a space-fractional wave equation}

\author[1]{John P. Hollkamp}

\affil[1]{Department of Mechanical Engineering, Ray W. Herrick Laboratories, Purdue University, West Lafayette,
IN 47907
}

\author[2]{Mihir Sen}
\affil[2]{Department of Aerospace and Mechanical Engineering, University of Notre Dame, Notre Dame,
IN 46556
}
\author[1]{Fabio Semperlotti}

\date{}

\maketitle

\begin{abstract}
This study explores the use of fractional calculus as a possible tool to model wave propagation in complex, heterogeneous media. While the approach presented could be applied to predict field transport in a variety of inhomogeneous systems, we illustrate the methodology by focusing on elastic wave propagation in a one-dimensional periodic rod. The governing equations describing the wave propagation problem in inhomogeneous systems typically consist of partial differential equations with spatially varying coefficients. Even for very simple systems, these models require numerical solutions which are computationally expensive and do not provide the valuable insights associated with closed-form solutions. We will show that fractional calculus can provide a powerful approach to develop comprehensive mathematical models of inhomogeneous systems that can effectively be regarded as homogenized models. Although at first glance the mathematics might appear more complex, these fractional order models can allow the derivation of closed-form analytical solutions that provide excellent estimations of the systems' dynamic responses. Equally important, these solutions are valid in a frequency range that goes largely beyond the well-known homogenization limit of traditional integer order approaches, therefore providing a possible route to high-frequency homogenization. More specifically, this study focuses on the analyses of the dispersion and propagation properties of a periodic medium under single-tone harmonic excitation and illustrates the methodology to obtain a space-fractional wave equation capable of capturing the behavior of the physical system. The fractional wave equation and its analytical solution are compared with numerical results obtained via a traditional finite element method in order to assess their validity and evaluate their performance. It is found that the resulting fractional differential models are, in their most general form, of complex and frequency-dependent order.

\end{abstract}

\section{Introduction} \label{sec:intro}

The rapid growth of fields such as composite and architectured materials (often referred to as metamaterials), along with continuing advancements in design and fabrication procedures have led to the synthesis of complex systems and structures having intricate material distributions and non-trivial geometries. Such systems include, but are not limited to, fractal microstructures \cite{Tarasov,Tarasov2}, highly anisotropic structures \cite{Meerschaert,Ponge, Torrent}, and porous materials \cite{Hornung,Fellah2}. The physical response of these systems is often characterized by a combination of complex phenomena including, for example, nonlocal interactions, memory effects, and multi-scale response in either time or space. In recent years, experimental investigations have made it increasingly evident that an overwhelming majority of physical phenomena dealing with both static and dynamic response of complex systems exhibit a combination of these mechanisms. The difficulty of achieving a proper mathematical representation of their response is a direct consequence of the outstanding complexity and multifaceted nature of the underlying physical mechanisms.

These are some of the reasons that have motivated, over many decades, the development of numerical approaches in order to tackle complex problems in the physics of continuous systems. A popular example is represented by the class of discretization methods (e.g. finite differences or finite elements). Although these approaches have proven to be extremely powerful and accurate for many different applications, they are subject to an implicit trade-off between accuracy and computational time which is controlled by the level of discretization in either the temporal or the spatial domains. These discretization methods break down quickly in presence of multiscale inhomogeneities (e.g. complex microstructures, porous or fractal media, hybrid field transport properties) that require high levels of discretization largely beyond computational capabilities. Equally important, these methodologies allow only a fully numerical approach therefore providing a limited insight on the effect of different system parameters and requiring to reevaluate the solution following every modification. As a result, there is a need in computational mechanics for methodologies that, on one side, are capable of reducing the order of the system while maintaining the solution's accuracy and, on the other, lead to compact analytical formulations paving the way to closed-form solutions of complex systems. 

Over the past several decades, various methodologies were explored spanning areas like model order reduction \cite{Besselink,deKlerk,Schilders}, perturbation methods \cite{Nayfeh}, and homogenization techniques \cite{Manevitch,Qin,Yu}. In the field of wave propagation in inhomogeneous media, the homogenization techniques have been some of the most popular and effective approaches to achieve both numerical and, sometimes, analytical solutions of complex media. Homogenization techniques allow modeling the macroscopic response of a inhomogeneous medium based on the properties of its microscopic constituents \cite{Babuska}. In essence, these techniques eliminate the spatial dependence typical of inhomogeneous materials by replacing the material properties with constant (or, eventually, frequency dependent) values. The result is a homogeneous medium exhibiting an equivalent response compared to the original material. The resulting mechanical properties of these homogenized models are known as effective properties and are determined via different mathematical techniques. In certain limited cases, homogenized models can also lead to closed-form analytical solutions, thus avoiding reliance on cumbersome numerical methods. While homogenization has served as an effective way to deal with media possessing complicated microstructures, homogenization techniques are typically restricted by key limitations which include dependency on \textit{ad hoc} assumptions and limited range of validity; the latter probably being the most restricting condition. In a context of dynamic simulations, the homogenized models are representative of the actual heterogeneous medium only within a selected wavelength range, the so-called long wavelength regime. Although the actual bounds of this regime cannot be rigorously defined and depend strongly on the specific nature of the material (e.g. on the presence of resonating elements), dynamic homogenized models are typically considered a good approximation when the wavelength is at least twice the unit cell size. A different aspect, still related to this intrinsic limit of classical homogenization techniques, concerns their inability to model the response of periodic systems within band gaps. Considering that the occurrence of frequency band gaps is mostly a manifestation of local effects (either Bragg scattering or local resonances), it is not surprising that homogenized models which conceptually rely on the global response of the medium break down within these frequency ranges. It is to target this specific limitation that, in recent years, researchers have been developing the so-called high-frequency homogenization techniques \cite{Antonakakis1,Antonakakis2}.

In this paper, we develop the groundwork for what can be effectively considered a homogenization technique for periodic media based on fractional calculus. This approach could potentially address two key shortcomings of current methods: 1) the inability to find closed form solutions capturing the system dynamics, and 2) the breakdown of standard homogenization methods within the band gap regimes. In particular, this study explores the possible use and the corresponding accuracy of space-fractional partial differential equations to model wave propagation in one-dimensional periodic media. From a general perspective, we will show how a space-fractional wave equation can be used in place of a traditional integer-order wave equation with spatially-dependent coefficients in order to obtain analytical solutions that well represent the wave propagation in the medium. 

The field of fractional calculus has only recently seen applications in engineering even though the purely mathematical study of integrals and derivatives of non-integer order has existed since the 17th century. Engineering areas that have successfully leveraged fractional calculus include viscoelasticity \cite{Torvik,Wharmby,Achar1}, fractals \cite{Carpinteribook,Carpinteri,Ostoja}, vibration control \cite{DiMatteo1,DiMatteo2}, damping \cite{Achar2,Ryabov,Tofighi}, systems exhibiting memory \cite{Tarasova,Wang}, and acoustic wave propagation in complex (mostly porous) media \cite{Fellah,Casasanta,Tarasov}. The reader is referred to \cite{Podlubny,Herrmann,Diethelm} for detailed reviews on the fundamentals of fractional calculus. In Appendix \ref{App_A}, we also provide a short summary of the basic definitions of fractional derivatives and of the corresponding transforms. Note that fractional derivatives in time are known to physically represent damping and dissipation that occur in lossy or viscoelastic materials. On the other hand, fractional derivatives in space are indicative of attenuation in systems that potentially are still conservative. That is, space fractional derivatives are ideal tools to capture frequency band gaps in which attenuation is due to multiple back scattering and not to energy dissipation. 

It is anticipated that the methodology discussed in this paper produces fractional differential models of complex and frequency-dependent order. The mathematics of complex fractional derivatives is an active area of mathematical research that is still being developed. Authors such as Love \cite{Love}, Ortigueira \cite{Ortigueira}, Ross \cite{Ross}, Andriambololona \cite{Andriambololona}, and Valerio \cite{ValerioJA} have worked on some aspects of complex fractional derivatives. However, applications involving complex fractional orders are very limited. Authors including Atanackovic \cite{Atanackovi}, Makris \cite{Makris}, and Hollkamp \cite{Hollkamp} have used complex fractional calculus to study various engineering systems. Despite the use of complex order fractional operators in these papers, the actual physical significance of a complex fractional order derivative is still not completely evident and continues to garner attention today. Makris provides a discussion linking complex-order derivatives to the modulation of both the amplitude and phase of harmonic components of a function. He shows that an ``\textit{important difference between real-valued and complex-valued time derivatives is that phase modulation in the latter case is frequency dependent whereas in the former is not}" \cite{Makris}. This statement, as well as other physical meanings of a complex order fractional derivative, will be further analyzed.

This paper extends the methodology developed in \cite{Hollkamp} for discrete systems to modeling continuous systems. In \cite{Hollkamp}, the authors developed fractional order differential models to represent the dynamic response of non-homogeneous discrete systems and to achieve an accurate model order reduction methodology. By extending this methodology to non-homogeneous continuous systems we obtain an approach that can be interpreted as a simplified fractional homogenization technique applicable to dynamic problems. 

The remainder of the paper is structured as follows: we first introduce the system under investigation, that is a bi-material periodic rod, and outline the general procedure. Then, we focus on the mathematical details necessary to obtain the fractional-order homogenized model of the continuum. Next, the analytical solution to the fractional wave equation is obtained by exploring different approaches. Finally, we present a comparison between the performance of the fractional model and a traditional finite element solution.

\section{One-dimensional inhomogeneous periodic systems: synthesis of fractional order models} \label{sec:description}

\begin{figure}[ht] 
  \begin{center}
    \centerline{\includegraphics[scale=0.65,angle=0]{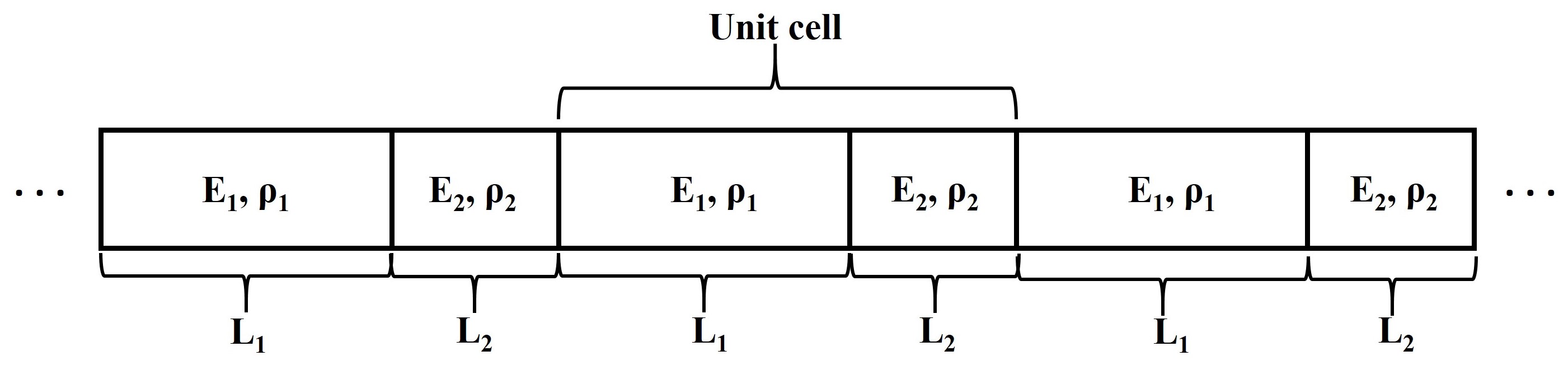}}
    \caption{Drawing of periodic, bi-material rod.}
    \label{fig:rod}
  \end{center}
\end{figure}

The main objective of this work is to develop a methodology to synthesize a space-fractional order model capable of capturing the dynamics of a periodic one-dimensional waveguide while offering a route to closed-form analytical solutions. In an effort to develop the building blocks of this methodology and to perform a preliminary evaluation of its performance, we selected a benchmark system of a one-dimensional rod made of two periodically alternating materials. The system is schematically illustrated in Fig.~\ref{fig:rod} and represents a solid rod whose Young's modulus and density vary periodically and in a step-like fashion between $[E_1,\rho_1]$ and $[E_2,\rho_2]$. The unit cell of this periodic structure is also indicated in Fig.~\ref{fig:rod}. This system is a classical example of a one-dimensional elastic metamaterial. In the following calculations, we assumed the two materials to be aluminum ($E_1=70$ GPa and $\rho_1 = 2700$ kg/m$^3$) and brass ($E_2=110$ GPa and $\rho_1 = 8100$ kg/m$^3$) as well as $L_1=L_2=1$ m.

The equation of motion for the rod in Fig.~\ref{fig:rod} is

\begin{eqnarray} \label{eq:EOM}
\frac{\partial}{\partial x} \bigg[ E(x)A \frac{\partial u}{\partial x}\bigg] = \rho(x)A \frac{\partial^2 u}{\partial t^2}
\end{eqnarray}

\noindent where $x$ is the spatial coordinate, $t$ is the time coordinate, $E$ is Young's modulus, $A$ is the cross-sectional area, $\rho$ is the density, and $u$ is the particle displacement. In this work, we will consider $A$ to be constant. Note that the analytical solution of Eq.~\eqref{eq:EOM} is known only for very specific variations of $E$, $\rho$, and $A$ \cite{Graff}. 

\subsection{General approach} \label{sec:method}

A classical approach often used to derive fractional differential equations capable of describing the fractional dynamics of a continuous system is to start with the integer-order equations of motion, apply a transformation (such as Laplace or Fourier transform), solve the equation in the transformed domain with the proper assumptions and boundary conditions, and then take an inverse transform to obtain an equivalent fractional derivative in the time-space domain \cite{Torvik,Fellah}. This approach works well when the underlying constitutive relations have an intrinsic power-law dependence (e.g. for viscoelastic or fractal media). However, in this study we consider systems that are not intrinsically fractional, but that can be modeled as fractional in order to gain an analytical advantage.

For this reason, we employ a different approach in order to avoid certain complexities that arise when taking the transform of partial differential equation having variable coefficients. Consider the space-fractional partial differential equation, given by

\begin{eqnarray} \label{eq:EOM_frac}
\bar{c}^2\frac{\partial^\alpha u}{\partial x^\alpha} = \frac{\partial^2 u}{\partial t^2}
\end{eqnarray}

\noindent where $\alpha$ is the order of the fractional derivative and $\bar{c}$ is the fractional wave ``speed" {that is an equivalent wave velocity having dimensions $\frac{m^{\alpha/2}}{s}$. In this work, we use a left-handed Caputo derivative (see \S \ref{App_A} for more detail) because it allows a more direct treatment of the boundary conditions and of the half space. Our approach relies on the assumption that Eqs.~\eqref{eq:EOM} and \eqref{eq:EOM_frac} should have equivalent characteristic equations. This is justified because, from a physical point of view, these characteristic equations represent the dispersion of the system which must be the same in both approaches.
More specifically, we convert both Eqs.~\eqref{eq:EOM} and \eqref{eq:EOM_frac} from the space-time domain to the wavenumber-frequency domain, allowing us to obtain the dispersion relations and set them equal to each other. Then, we solve for the fractional order $\alpha$ that satisfies this equality. For clarity, we label this step of the analysis as phase one. The second phase of the analysis involves finding the analytical solution to Eq.~\eqref{eq:EOM_frac} with proper boundary conditions.
Once the analytical closed-form solution of the fractional wave equation for the periodic system has been obtained, we will assess the validity of the approach and the accuracy of the solution by comparing the response of the system at steady state with a reference solution obtained numerically by the finite element method.

The following sections of the paper will detail the three phases described above. Note that, although we focus on a periodic system, most of the methodology could be extended to any general one-dimensional inhomogeneous system and, with proper modifications, also to higher-dimensional systems.

\subsection{Dispersion analysis of the periodic rod}

The first phase of our method involves the calculation of the order $\alpha$ of the fractional wave equation such that the dispersion behavior matches that of the corresponding integer order model. To this extent, we first calculate the dispersion relation by using the integer order wave equation solved via a spectral method, and then we repeat this calculation for a space-fractional wave equation. Then, we use these two relations to identify a closed form solution that allows the two systems  to have identical dispersive behavior.

\subsubsection{Integer-order wave equation: dispersion calculation via spectral elements} \label{sec:dispersionSE}

Starting from the integer-order wave equation Eq.~\eqref{eq:EOM}, the dispersion relation can be found by using the well-established spectral elements approach and Floquet theory as described in \cite{Doyle}. The dispersion relation is found to be 

\begin{eqnarray} \label{eq:mu}
\textrm{cos}(\mu L) = \Big[\textrm{cos}\Big(\omega \frac{L_1}{c_1}\Big) \textrm{cos}\Big(\omega \frac{L_2}{c_2}\Big) - \frac{1}{2}\Psi \textrm{sin}\Big(\omega \frac{L_1}{c_1}\Big) \textrm{sin}\Big(\omega \frac{L_2}{c_2}\Big)\Big]
\end{eqnarray}

\noindent where 
\begin{eqnarray} \label{eq:Psi}
\Psi = \frac{A_1 E_1 c_2}{A_2 E_2 c_1}+\frac{A_2 E_2 c_1}{A_1 E_1 c_2}
\end{eqnarray}

\begin{eqnarray} 
L = L_1 + L_2
\end{eqnarray}

\noindent and $c_1$ is the wave speed in material 1, $c_2$ is the wave speed in material 2, and $\mu$ is the wavenumber of the entire bi-material, periodic rod configuration.

A plot of Eq.~\eqref{eq:mu} is provided in Fig.~\ref{fig:dispersion}. The plot shows both Re$(\mu)$ and Im$(\mu)$ versus frequency. Note that when the Im$(\mu)$ is nonzero, the elastic wave is evanescent. These regions corresponds to the so-called \textit{band gaps}, that is frequency ranges in which waves attenuate quickly or, equivalently, wave propagation through the system is not allowed. These zones are highlighted by gray areas in Fig.~\ref{fig:dispersion}.

\begin{figure}[h!] 
  \begin{center}
    \centerline{\includegraphics[scale=0.65,angle=0]{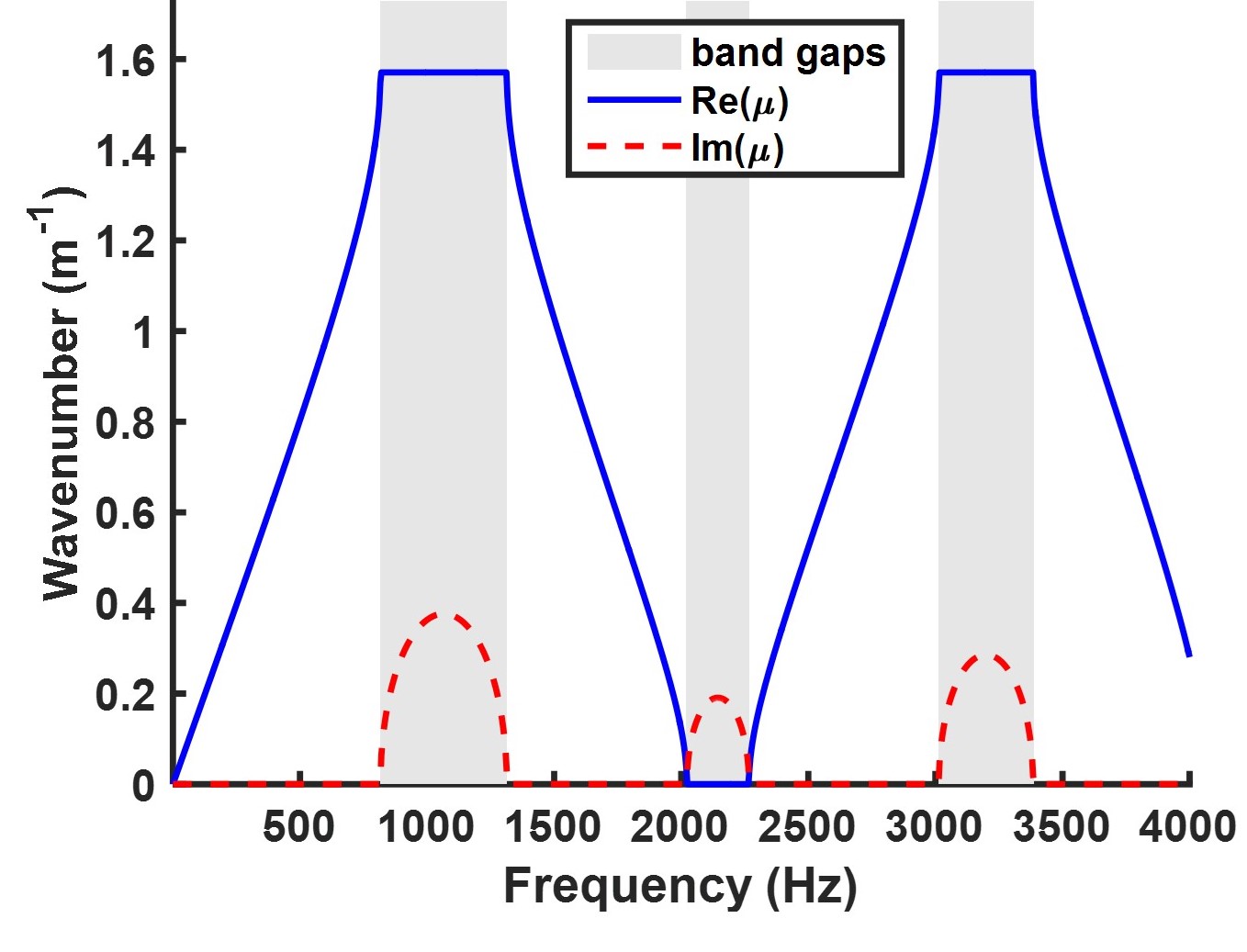}}
    \caption{Plot of the dispersion of the aluminum-brass periodic rod with band gaps depicted. The curve is wrapped over the first Brillouin zone.}
    \label{fig:dispersion}
  \end{center}
\end{figure}

\subsubsection{Fractional-order wave equation: dispersion calculation} \label{sec:dispersionFWE}

The general form of the space-fractional wave equation was previously defined in Eq.~\eqref{eq:EOM_frac}. The fractional derivative should be intended as a left handed Caputo derivative (see Appendix \ref{App_A}) which depends on the value of the lower bound of the integral. If the fractional derivative in Eq.~\eqref{eq:EOM_frac} has a lower bound of $-\infty$, then 

\begin{eqnarray} \label{eq:fracexp}
\leftidx{_{-\infty}^C}{D}{_x^\alpha}  [\textrm{e}^{bx}] = b^\alpha \textrm{e}^{bx}
\end{eqnarray}

\noindent where $\leftidx{_{-\infty}^C}{D}{_x^\alpha}$ is the operational notation for a Caputo fractional derivative of order $\alpha$ with lower bound $-\infty$ \cite{Meerschaert}. Then, the solution of the fractional differential Eq.~\eqref{eq:EOM_frac} is an exponential function \cite{Meerschaert} where the general form is

\begin{eqnarray} \label{eq:fracexpsol}
u(x,t)=Ae^{\textrm{i}(\omega t-kx)} + Be^{\textrm{i}(\omega t+kx)}.
\end{eqnarray}

This solution corresponds to forward and backward propagating waves, respectively. Substituting the first term of the right hand side of Eq.~\eqref{eq:fracexpsol} into Eq.~\eqref{eq:EOM_frac} and using Eq.~\eqref{eq:fracexp} leads to

\begin{eqnarray} \label{eq:fracdisp}
k = \textrm{i}\Big(-\frac{\omega^2}{\bar{c}^2}\Big)^{1/\alpha}.
\end{eqnarray}

\noindent The substitution of the second term of the right hand side of Eq.~\eqref{eq:fracexpsol} into Eq.~\eqref{eq:EOM_frac} would have yielded the negative value of Eq.~\eqref{eq:fracdisp}. 

If the fractional derivative of Eq.~\eqref{eq:EOM_frac} had a lower terminal of 0, then the exponential solution given by Eq.~\eqref{eq:fracexpsol} would no longer be a valid solution to Eq.~\eqref{eq:EOM_frac}. Instead, the solution can be stated in the form of the following ansatz (see also \S \ref{sec:MLsol}) as

\begin{eqnarray} \label{eq:fracanz}
u(x,t) = \textrm{e}^{\textrm{i}\omega t}[A\textrm{E}_{\alpha,1}((-\textrm{i}k x)^\alpha) + Bx\textrm{E}_{\alpha,2}((-\textrm{i}k x)^\alpha)]
\end{eqnarray}

\noindent where E$(\cdot)$ is the Mittag-Leffler function \cite{Podlubny, Gorenflo}, $A=u(0,t)$ and $B=\frac{du(0,t)}{dx}$. The Mittag-Leffler function is defined as 

\begin{eqnarray} \label{eq:Mittag}
E_{\alpha,\beta}(z) = \sum_{n=0}^{\infty} \frac{z^n}{\Gamma(\alpha n+\beta)}
\end{eqnarray}

\noindent where $\Gamma(\cdot)$ is the gamma function \cite{Gorenflo}, [$\alpha$,$\beta$] $\in \mathbb{C}$ are constants, and $z \in \mathbb{C}$. 

It is straightforward to verify that the solution given by Eq.~\eqref{eq:fracanz} satisfies Eq.~\eqref{eq:EOM_frac} by utilizing the following three formulas. First, the left-handed Riemann-Liouville derivative of a Mittag-Leffler function \cite{Podlubny} is

\begin{eqnarray} \label{eq:derivML}
\leftidx{_0^{RL}}{D}{_x^\gamma}  [x^{\beta-1}E_{\alpha,\beta}(\lambda x^\alpha)] = x^{\beta-\gamma-1}E_{\alpha,\beta-\gamma}(\lambda x^\alpha)
\end{eqnarray}

\noindent where $\gamma$ is the order of the derivative. The analytical relation between the Riemann-Liouville and Caputo derivatives is

\begin{eqnarray} \label{eq:RLtoC}
\leftidx{_{0}^C}{D}{_x^\gamma} u(x) = \leftidx{_0^{RL}}{D}{_x^\gamma}  u(x) - \sum_{q=0}^{n-1} \frac{d^q u}{dx^q}(0) \frac{x^{-\gamma+q}}{\Gamma(-\gamma+q+1)}
\end{eqnarray}

\noindent where $n = \ceil{\gamma}$ \cite{Sousa}. Finally, an important recurrence relation of the Mittag-Leffler function is \cite{Gorenflo}

\begin{eqnarray} \label{eq:recurr}
E_{\alpha,\beta}(z) = zE_{\alpha,\alpha+\beta}(z) + \frac{1}{\Gamma(\beta)}.
\end{eqnarray}

\noindent Substituting Eq.~\eqref{eq:fracanz} into Eq.~\eqref{eq:EOM_frac} and using Eq.~\eqref{eq:derivML} and Eq.~\eqref{eq:RLtoC} yields

\begin{eqnarray} 
\textrm{e}^{\textrm{i}\omega t}\bigg[ Ax^{-\alpha} E_{\alpha,1-\alpha}\Big((-\textrm{i}k)^\alpha x^\alpha\Big)+B x^{1-\alpha}  E_{\alpha,2-\alpha}\Big((-\textrm{i}k)^\alpha x^\alpha\Big) +  -A \frac{x^{-\alpha}}{\Gamma(1-\alpha)} -B\frac{x^{1-\alpha}}{\Gamma(2-\alpha)}   \bigg] = -\frac{\omega^2}{\bar{c}^2}u.
\end{eqnarray}

\noindent Using Eq.~\eqref{eq:recurr} and some algebraic work, this reduces to

\begin{eqnarray} \label{eq:fracdisp3}
k = \textrm{i}\Big(-\frac{\omega^2}{\bar{c}^2}\Big)^{1/\alpha}.
\end{eqnarray}

\noindent Eq.~\eqref{eq:fracdisp3} is exactly the same as Eq.~\eqref{eq:fracdisp}. Therefore, both the exponential and Mittag-Leffler solutions produce the same dispersion relationship under the proper assumptions for the bounds of the differintegral operator.

\subsection{Identification of the fractional order} \label{sec:matching}

In order to identify the fractional order $\alpha$ which guarantees that both wave equations yield the same dispersion, we set $\mu$ from Eq.~\eqref{eq:mu} equal to $k$ from Eq.~\eqref{eq:fracdisp3}. To determine $\mu$ from Eq.~\eqref{eq:mu}, an inverse cosine must be taken which yields a non-unique solution. The theory used to derive Eq.~\eqref{eq:mu} used a solution form $\bar{A}e^{-i \mu x}$. For the solution to exponentially decay in the band gaps, the imaginary part of $\mu$ must be negative. As it turns out, in order to meet this requirement, the left hand side of Eq.~\eqref{eq:mu} should be cos$(- \mu L)$. Thus, to solve for the fractional order $\alpha$, we set $\mu$ from the cos$(- \mu L)$ solution from Eq.~\eqref{eq:mu} equal to $k$ from Eq.~\eqref{eq:fracdisp3}. The procedure results in an equation given by

\begin{eqnarray} \label{eq:transcend}
\frac{1}{L} \textrm{cos}^{-1} \Big[\textrm{cos}\Big(\omega \frac{L_1}{c_1}\Big) \textrm{cos}\Big(\omega \frac{L_2}{c_2}\Big) - \frac{1}{2}\Psi \textrm{sin}\Big(\omega \frac{L_1}{c_1}\Big) \textrm{sin}\Big(\omega \frac{L_2}{c_2}\Big)\Big] = -\textrm{i}\Big(-\frac{\omega^2}{\bar{c}^2}\Big)^{1/\alpha}.
\end{eqnarray}

Eq.~\eqref{eq:transcend} can be numerically solved for $\alpha$. We note that when passing from an integer order differential equation with variable coefficients to a fractional-order equation having constant coefficients some assumptions must be made on the approach to convert the wave speed $c$ into the equivalent fractional wave speed $\bar{c}$. Before solving Eq.~\eqref{eq:transcend} for $\alpha$, we discuss some important implications in the selection of $\bar{c}$.

\subsubsection{Equivalent wave speed: constant speed assumption} \label{sec:conC}

A reasonable assumption for the value of the equivalent wave speed is to take $\bar{c}$ to be a constant value provided by the conventional homogenization approach in the long wavelength limit. 
According to \cite{Mei,Berryman}, the effective modulus $\bar{E}$ of the periodic rod can be obtained by the inverse rule of mixtures as

\begin{eqnarray}
\frac{1}{\bar{E}} = \frac{L_1/(L_1+L_2)}{E_1}+ \frac{L_2/(L_1+L_2)}{E_2}
\end{eqnarray}

\noindent while the effective density $\bar{\rho}$ is given by a weighted sum as

\begin{eqnarray}
\bar{\rho} = \rho_1 L_1/(L_1+L_2)  + \rho_2 L_2/(L_1+L_2). 
\end{eqnarray}

\noindent The effective wave speed is then defined to be

\begin{eqnarray} \label{eq:cbar}
\bar{c} = Q \sqrt{\frac{\bar{E}}{\bar{\rho}}}
\end{eqnarray}

\noindent where $Q$ is a dimensional multiplicative factor that ensures units consistency. Here, $[Q]=[m^{\alpha/2 -1}]$.
Setting $\bar{c}$ according to Eq.~\eqref{eq:cbar} and substituting into Eq.~\eqref{eq:transcend}, we obtain the frequency-dependent order $\alpha$ to be

\begin{eqnarray} \label{eq:alpha}
\alpha = \frac{2\textrm{ln}(\omega/\bar{c})+\textrm{i}\pi}{\textrm{ln}(\mu)+\textrm{i}\pi/2}.
\end{eqnarray}

A plot of Eq.~\eqref{eq:alpha} for the periodic rod is shown in Fig.~\ref{fig:alpha}. There are a few interesting features in this plot that deserve some discussion. First, the fractional order is a complex number which indicates that the fractional equivalent (homogenized) model will be described by complex order differ-integral operators. As discussed in the introduction, a complex order derivative allows for the frequency-dependent modulation of both the phase and the amplitude \cite{Makris} of harmonic components, therefore allowing for virtually unrestricted matching of the dispersion relations. 

The complex fractional order undergoes some interesting transitions when spanning different frequency regimes.
In the low frequency range before the first band gap (typically considered as the range of validity for homogenization models under the long wavelength assumption), the value of $\alpha$ is actually purely real and equal to 2. This means that, within the long wavelength limit, the space-fractional wave model with a constant equivalent speed reduces to the classic integer-order homogenized model. 

Moving to higher frequencies within the first band gap, it is observed that Re$(\alpha)<2$ and Im$(\alpha)\ne 0$ except for few selected values of $\omega$. This behavior is consistent with the expected amplitude attenuation characteristic of a band gap and with the fact that space-fractional wave equations with $1<$ Re$(\alpha)<2$ correspond to spatially attenuated waves \cite{Achar2,Ryabov,Tofighi,Meerschaert}. As we move to band gaps in the higher frequency range, we observe that, typically, Re$(\alpha)<2$. Also, in the pass band, Re$(\alpha)\approx 2$. This is again consistent with the physical nature of the problem. For a few frequency values, we see that Re$(\alpha)>$2, which could raise some potential stability issues. This has motivated us to also explore different directions to estimate $\bar{c}$. 

\begin{figure}[h!] 
  \begin{center}
    \centerline{\includegraphics[scale=0.65,angle=0]{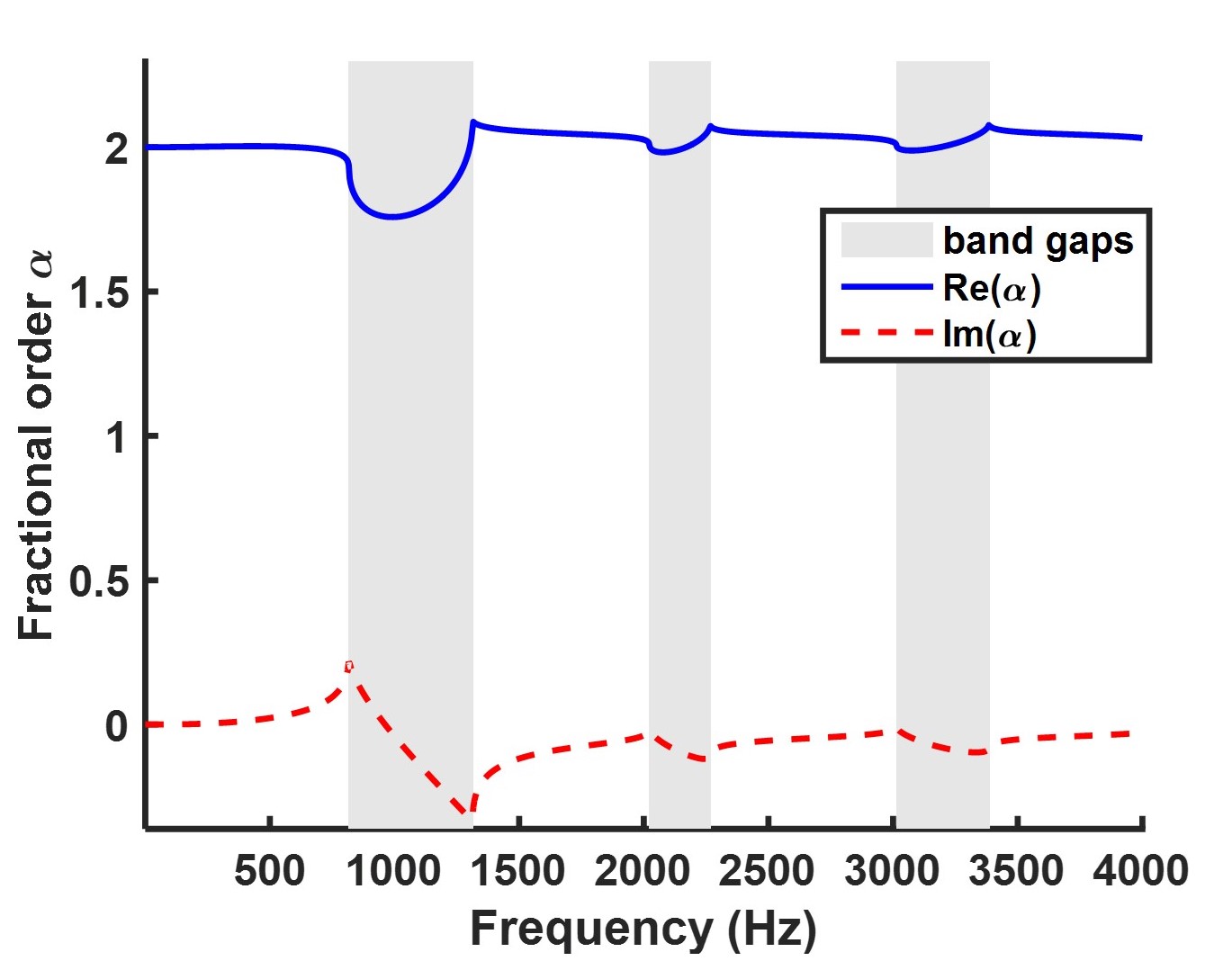}}
    \caption{The complex, frequency-dependent fractional order $\alpha$ for the aluminum-brass example.}
    \label{fig:alpha}
  \end{center}
\end{figure}

\subsubsection{Equivalent wave speed: frequency-dependent speed assumption} \label{sec:varC}

Due to the above mentioned stability issues, we developed another approach for $\bar{c}$ where both $\alpha$ and $\bar{c}$ depend on the frequency $\omega$. Recall the values of $\mu(\omega)$ that is given by Fig.~\ref{fig:dispersion} and the left hand side of Eq.~\eqref{eq:transcend}. According to Meerschaert \cite{Meerschaert}, the wavenumber can be written as $\mu = \omega/c(\omega) +$ i$\eta$ where $c(\omega)$ is a frequency dependent wave speed and $\eta$ is an attenuation factor. From this, we can say that the speed $\bar{c}(\omega)$ is

\begin{eqnarray} \label{eq:cbarfreq}
\bar{c} = Q \frac{\omega}{\textrm{Re}(\mu)}.
\end{eqnarray}

A plot of $\bar{c}(\omega)$ for the periodic rod is given in Fig.~\ref{fig:varC}a. For very low frequencies in the first pass band the value of $\bar{c}$ is equal to the value given by Eq.~\eqref{eq:cbar}. The trends of $\bar{c}$ in the band gaps are linear while the curves in the pass bands vary smoothly.

\begin{figure}[h!] 
  \begin{center}
    \centerline{\includegraphics[scale=0.65,angle=0]{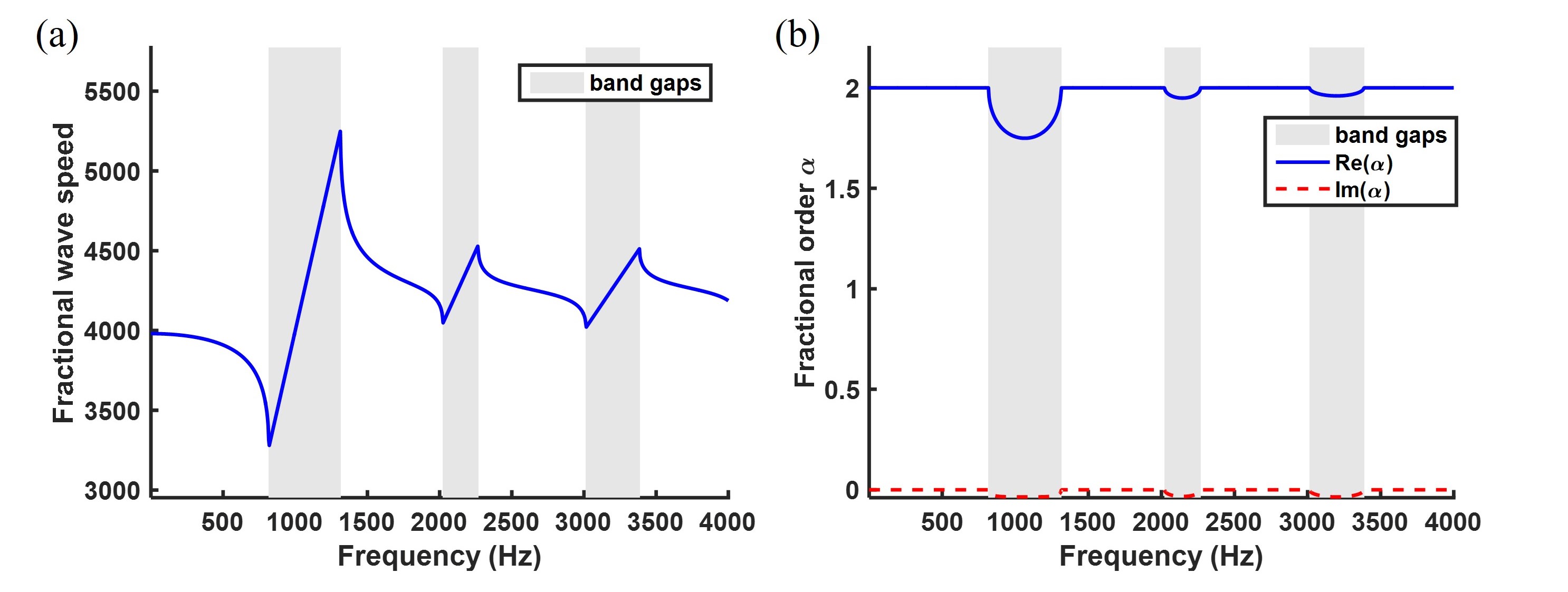}}
    \caption{a) Plot of the fractional wave speed $\bar{c}$ as a function of frequency. b) The corresponding fractional order $\alpha$ whose real part is never greater than 2.}
    \label{fig:varC}
  \end{center}
\end{figure}

Using this $\bar{c}(\omega)$, Eq.~\eqref{eq:transcend} can be solved for $\alpha$ as a function of frequency. Substituting Eq.~\eqref{eq:cbarfreq} into Eq.~\eqref{eq:transcend} and performing some algebraic simplifications results in

\begin{eqnarray} \label{eq:alpha2}
\alpha = \frac{2\textrm{ln}(|\textrm{Re}(\mu)|)+\textrm{i}\pi}{\textrm{ln}(\mu)+\textrm{i}\pi/2}.
\end{eqnarray}

The plot of Eq.~\ref{eq:alpha2} for the periodic rod is given in Fig.~\ref{fig:varC}b. In contrast with the previous method, now Re$(\alpha)=2$ for all the frequencies in pass bands. This is again consistent with the physics since an order of two of the operator is indicative of propagating waves. It is also not surprising because we have reinstated some degree of variability in the coefficients of the equation, albeit the variability does not occur as a function of space; rather, it is dependent upon frequency. In the band gaps, Re$(\alpha)<2$ while Im$(\alpha)<0$. The fact that Re$(\alpha)<2$ indicates that the wave propagation decays, consistent with the fact that this frequency is located in a band gap. Again, the meaning of the imaginary part being non-zero is difficult to pinpoint exactly, but as we will see in the results, the imaginary part affects the phase of the results. In addition, we see that Re$(\alpha) \leq 2$ at all frequencies, avoiding some of the stability issues that resulted from the former approach of using the same value of $\bar{c}$ for all frequencies. When using this approach in phase 2, we will consider a harmonic input of frequency $\Omega$. The corresponding values of $\bar{c}$ of $\alpha$ used in Eq.~\eqref{eq:EOM_frac} will be found by using the results in Fig.~\ref{fig:varC} at a frequency $\Omega$.

\section{Closed-form analytical solution}

The previous sections presented a methodology to determine the order $\alpha(\omega)$ of the space fractional operator that guarantees an equivalent dispersion behavior with respect to an integer-order wave equation for the same periodic system. In this section, we explore possible closed-form analytical solutions corresponding to the forced response of the periodic rod under harmonic excitation conditions. 

As a reminder, it was previously determined that the solution to Eq.~\eqref{eq:EOM_frac} depends on the value of the lower bound of the fractional derivative. When the lower bound is $-\infty$, the solution is based on an exponential kernel. When the lower bound is zero, the solution is based on Mittag-Leffler functions. We will consider both solutions.

\subsection{The exponential kernel} \label{sec:expsol}

The exponential solution was given by Eq.~\eqref{eq:fracexpsol}. Since the exponential solution requires a lower bound of $-\infty$, the exponential solution is applicable for an infinite or semi-infinite rod. Consider a semi-infinite homogenized model that is subjected to an harmonic input of frequency $\Omega$ and only consists of waves propagating to the right; that is, in Eq.~\eqref{eq:fracexpsol}, set $B$=0 and $\omega$=$\Omega$. This yields

\begin{eqnarray} \label{eq:Meersolright}
u(x,t)=Ae^{\textrm{i}(\Omega t-kx)}.
\end{eqnarray}

\noindent Using the dispersion relation from Eq.~\eqref{eq:fracdisp}, Eq.~\eqref{eq:Meersolright} can be re-written as

\begin{eqnarray} \label{eq:entireMeersol}
u(x,t) = A\textrm{e}^{\textrm{i}\Omega t} \textrm{e}^{ (-\Omega^2/\bar{c}^2)^\frac{1}{\alpha}x}.
\end{eqnarray}

\noindent Recall that the value of $\alpha$ was previously obtained. Consider only the spatial term of Eq.~\eqref{eq:entireMeersol}, given again here as

\begin{eqnarray} \label{eq:Meerspace}
\phi(x) = \textrm{e}^{ (-\Omega^2/\bar{c}^2)^\frac{1}{\alpha}x}.
\end{eqnarray}

\noindent Let us analyze the stability of this solution. Define $\lambda=\frac{\Omega^2}{\bar{c}^2}$ and $\alpha=\zeta+\epsilon\textrm{i}$; thus, Eq.~\eqref{eq:Meerspace} can be written as

\begin{eqnarray} \label{eq:Meerspace2}
\phi(x) = \textrm{Exp}\bigg[{ (-\lambda)^\frac{\zeta-\epsilon\textrm{i}}{\zeta^2+\epsilon^2}x}\bigg].
\end{eqnarray}

Note that for lengthy mathematical expressions we will use both the notations $e^{(\cdot)}$ and Exp${(\cdot)}$ to indicate the exponential function. Let $a=\frac{\zeta}{\zeta^2+\epsilon^2}$ and $b=\frac{\epsilon}{\zeta^2+\epsilon^2}$. After some algebraic work, the solution can be written as

\begin{eqnarray} 
\phi(x) = \textrm{Exp}\bigg[\lambda^a \textrm{e}^{b\pi} \textrm{cos}(a\pi-b\textrm{ln}(\lambda)) x \bigg] \bigg[\textrm{cos}\bigg( \lambda^a \textrm{e}^{b\pi} \textrm{sin}(a\pi-b\textrm{ln}(\lambda)) x\bigg)+\textrm{i}\textrm{sin}\bigg( \lambda^a \textrm{e}^{b\pi} \textrm{sin}(a\pi-b\textrm{ln}(\lambda)) x\bigg) \bigg].
\end{eqnarray}

\noindent The real part of Eq.~\eqref{eq:Meerspace} is

\begin{eqnarray} \label{eq:MeerspaceRe}
\textrm{Re}(\phi(x)) = \textrm{Exp}\bigg[\lambda^a \textrm{e}^{b\pi} \textrm{cos}(a\pi-b\textrm{ln}(\lambda)) x \bigg] \bigg[\textrm{cos}\bigg( \lambda^a \textrm{e}^{b\pi} \textrm{sin}(a\pi-b\textrm{ln}(\lambda)) x\bigg) \bigg]
\end{eqnarray}

\noindent while the imaginary part is

\begin{eqnarray} \label{eq:MeerspaceIm}
\textrm{Im}(\phi(x)) = \textrm{Exp}\bigg[\lambda^a \textrm{e}^{b\pi} \textrm{cos}(a\pi-b\textrm{ln}(\lambda)) x \bigg] \bigg[\textrm{sin}\bigg( \lambda^a \textrm{e}^{b\pi} \textrm{sin}(a\pi-b\textrm{ln}(\lambda)) x\bigg) \bigg].
\end{eqnarray}

The trigonometric terms on the right hand side of Eqs.~\eqref{eq:MeerspaceRe} and \eqref{eq:MeerspaceIm} are oscillatory and thus are stable. However, the exponential terms will determine if the solution exponentially decays or grows. The terms $\lambda^a$ and e$^b\pi$ are always positive. Thus, the term that determines the stability is cos$(a\pi-b\textrm{ln}(\lambda))$. Cosine is negative if its argument is between $m\frac{\pi}{2}$ and $m\frac{3\pi}{2}$ where $m$ is an odd integer. As a result, Eq.~\eqref{eq:Meerspace} is stable if and only if

\begin{eqnarray} \label{eq:stabcond}
m\frac{\pi}{2} \leq \Big(a\pi-b\textrm{ln}(\lambda)\Big) \leq m\frac{3\pi}{2} & \text{where $m$ is an odd integer.}
\end{eqnarray}

If $a\pi-b\textrm{ln}(\lambda)= m\frac{\pi}{2}$ (where $m$ is odd), the exponential solution will be marginally stable - that is, the solution neither grows without bounds nor decays. If the equality part of Eq.~\eqref{eq:stabcond} is removed, the solution will be unconditionally stable - it will always decay. If Eq.~\eqref{eq:stabcond} is not satisfied, then the solution will grow exponentially. Since the values of $a$, $b$, and $\lambda$ all contribute to whether Eq.~\eqref{eq:stabcond} is satisfied, we cannot definitively conclude what the limiting values of $a$, $b$, and $\lambda$ are for stability. However, consider a purely real value of $\alpha$=2. This corresponds to $a$=$1/2$ and $b$ = 0. Thus, $a\pi-b\textrm{ln}(\lambda)$ = $\pi/2$, showing that the solution is marginally stable. Consider purely real orders greater than 2. The value of $b$ is still zero while $a<1/2$. Therefore, the solution will be unstable. However, if we use a complex order $\alpha$, it may be possible, depending on what the values of $b$ and $\lambda$ are, that Eq.~\eqref{eq:stabcond} can be satisfied even though $a<1/2$. It is obvious then that Im($\alpha$) contributes to the stability of the solution (in addition to the phase). In fact, the values of $\alpha(\omega)$ given in Fig.~\ref{fig:alpha} actually always satisfy Eq.~\eqref{eq:stabcond} even when Re$(\alpha)>2$. The potential stability issues for Re$(\alpha)>$2 can occur when the Mittag-Leffler kernel is implemented.

\subsection{The Mittag-Leffler kernel} \label{sec:MLsol}

Previously we assumed an ansatz for Eq.~\eqref{eq:fracanz} based on Mittag-Leffler (ML) functions. Here, we justify the use of this ansatz. To obtain the solution to Eq.~\eqref{eq:EOM_frac} when the Caputo fractional derivative has a lower terminal of zero, separation of variables and Laplace transforms are applied. Using separation of variables means that

\begin{eqnarray} \label{eq:SOV}
u(x,t)=\phi(x) q(t).
\end{eqnarray}

\noindent Substituting Eq.~\eqref{eq:SOV} into Eq.~\eqref{eq:EOM_frac}, and separating the time and space variables yields

\begin{eqnarray} \label{eq:separated}
\frac{\ddot{q}}{q} = \frac{\bar{c}^2}{\phi} \frac{\partial^\alpha \phi}{\partial x^\alpha} = -\tau^2
\end{eqnarray}

\noindent where $\tau$ is a positive constant (to guarantee stable oscillatory solutions). The temporal part of Eq.~\eqref{eq:separated} results in the classical second-order ordinary differential equation, which is

\begin{eqnarray} \label{eq:timeODE}
\frac{d^2 q}{dt^2}+\tau^2 q=0.
\end{eqnarray}

\noindent Recall the solution to Eq.~\eqref{eq:timeODE} is 

\begin{eqnarray} \label{eq:timesol}
q(t)=c_1 \textrm{e}^{\textrm{i}\tau t}+ c_2\textrm{e}^{-\textrm{i}\tau t}
\end{eqnarray}

\noindent where $c_1$ and $c_2$ are constants to be determined. Alternatively, the solution to Eq.~\eqref{eq:timeODE} could be written using a summation of cosine and sine terms. Now let $\xi$ = $\frac{\tau^2}{\bar{c}^2}$, allowing us to write the spatial part of Eq.~\eqref{eq:separated} as 

\begin{eqnarray} \label{eq:spaceODE}
\frac{d^\alpha \phi}{dx^\alpha}+\xi\phi = 0.
\end{eqnarray}

When 0 $< \alpha <$ 1, Eq.~\eqref{eq:spaceODE} is known as the fractional relaxation equation \cite{Podlubny,Herrmann}, while when 1 $< \alpha <$ 2, Eq.~\eqref{eq:spaceODE} is known as the fractional oscillation equation \cite{Podlubny,Herrmann}. Eq.~\eqref{eq:spaceODE} can be solved by using the Laplace transform (see Eq.~\eqref{LTC}) hence obtaining

\begin{eqnarray} \label{eq:C-trans}
\Phi (s) = A\frac{s^{\alpha-1}}{s^\alpha+\xi}+B\frac{s^{\alpha-2}}{s^\alpha + \xi}
\end{eqnarray}

\noindent where $s$ is the Laplace variable, $\Phi (s)$ is the Laplace transform of $\phi(x)$, $A$ = $\phi (0)$ and $B=\frac{d\phi (0)}{dx}$. The form of Eq.~\eqref{eq:C-trans} will result in Mittag-Leffler functions since the Laplace transform of the Mittag-Leffler function is \cite{Duan,Podlubny}

\begin{eqnarray} \label{eq:LTML}
\mathcal{L} \Big(z^{\beta-1}E_{\alpha,\beta}(\pm az^\alpha)\Big) =\frac{s^{\alpha-\beta}}{s^\alpha \mp a}. 
\end{eqnarray}

\noindent Using Eq.~\eqref{eq:LTML} to take the inverse Laplace transform of Eq.~\eqref{eq:C-trans} yields

\begin{eqnarray} \label{eq:spacesol}
\phi(x)=A E_{\alpha,1}(-\xi x^\alpha)+Bx E_{\alpha,2}(-\xi x^\alpha)
\end{eqnarray}

\noindent with $A$ = $\phi (0)$ and $B=\frac{d\phi (0)}{dx}$. Note that the form of Eq.~\eqref{eq:spacesol} is equivalent to the spatial part of Eq.~\eqref{eq:fracanz} (when $-\xi$=$(-\textrm{i}k)^\alpha$) hence confirming the ansatz. Furthermore, when $\alpha=2$, the exponential and Mittag-Leffler solutions are equivalent. See Appendix \ref{App_B} for more details.

\subsection{Boundary Conditions} \label{sec:Bc}

Consider an infinite homogenized rod having an assigned harmonic displacement at $x=0$ at a frequency $\Omega$, meaning that 

\begin{equation} \label{eq:harmBC}
u(0,t) = u_0 \textrm{e}^{\textrm{i} \Omega t}
\end{equation}  

\noindent where $u_0$ is a constant. The solution form using the exponential kernel is

\begin{eqnarray} \label{eq:fracexpsol2}
u(x,t)=Ae^{\textrm{i}(\Omega t-kx)} + Be^{\textrm{i}(\Omega t+kx)}.
\end{eqnarray}

We note that in such a configuration the forward and backward traveling waves from $x$=0 are equivalent, hence allowing the study of an infinite rod problem excited at one location to be accomplished by merely considering half of the domain (i.e. the semi-infinite rod). If the analysis is limited, for instance, to the positive x-axis, then $B$=0 and $A=u_o$. Clearly, the situation would reverse ($B=u_o$ and $A$=0) when considering the negative x-axis. 
Assuming the positive x-axis domain, the exponential solution for the semi-infinite fractional-order wave model subject to a harmonic displacement of amplitude $u_o$ and frequency $\Omega$ at $x$=0 is 

\begin{eqnarray} \label{eq:entiresolM}
u(x,t) = u_0\textrm{e}^{\textrm{i}\Omega t} \textrm{e}^{ (-\Omega^2/\bar{c}^2)^\frac{1}{\alpha}x}
\end{eqnarray}

\noindent where the value of $k$ was given by Eq.~\eqref{eq:fracdisp}. Now consider the semi-infinite rod using the Mittag-Leffler solution. The solution is given by multiplying Eq.~\eqref{eq:timesol} and \eqref{eq:spacesol}. Implementing Eq.~\eqref{eq:harmBC} requires that $a=u_0$, $\tau = \Omega$, $c_1$=1, and $c_2$=0. As a result, the product of Eq.~\eqref{eq:timesol} and \eqref{eq:spacesol} simplifies to

\begin{equation} \label{eq:solABC}
u(x,t) = e^{i \Omega t} \bigg[u_o E_{\alpha,1}\Big(-\frac{\Omega^2}{\bar{c}^2}x^\alpha\Big) + B x E_{\alpha,2}\Big(-\frac{\Omega^2}{\bar{c}^2}x^\alpha\Big)\bigg].
\end{equation}

Recall that from previous derivations $B=\frac{du(0,t)}{dx}$. To determine $B$, we use the known relationship between force and velocity - that is, mechanical impedance - at the location $x$=0, which is

\begin{equation} \label{eq:mechimp}
F = Zv
\end{equation}

\noindent where $F$ is the force, $v$ is the velocity, and $Z$ is the mechanical impedance. Considering that the displacement is known to be harmonic and has an amplitude of $u_o$, the velocity amplitude can be obtained by differentiation: $v$ = i$\Omega u_o$. The internal force is given by making use of the constitutive relationship for a rod: $F =  EA\frac{du}{dx}$. Hence, Eq.~\eqref{eq:mechimp} becomes

\begin{equation} 
EA\frac{du(0,t)}{dx} = \textrm{i}\Omega u_o Z.
\end{equation}

\noindent The mechanical impedance of the rod is $Z= \rho c A$, resulting in

\begin{equation} 
EA\frac{du(0,t)}{dx} = \textrm{i}\Omega u_o \rho c A.
\end{equation}

\noindent Recall that the wave speed $c$ can be written in terms of $E$ and $\rho$ as

\begin{equation} 
c = \sqrt{\frac{E}{\rho}}.
\end{equation}

\noindent Finally, $B$ is obtained as

\begin{equation} \label{eq:BMI}
\frac{du(0,t)}{dx} = B = \textrm{i}\Omega \frac{u_o}{c}.
\end{equation}

In Eq.~\eqref{eq:BMI}, the value of $c$ can be set to $\bar{c}$ according to one of the methodologies described in \S \ref{sec:conC} or \S \ref{sec:varC}. Contrary to the exponential solution, the Mittag-Leffler functions in Eq.~\eqref{eq:solABC} are unstable whenever Re$(\alpha)>2$ \cite{Diethelm}, no matter the value of the imaginary part. We will see the effects of this in the next section.

\section{Numerical assessment of the methodology} \label{sec:comparison}

In order to validate the proposed fractional model and its performance in the frame of the periodic rod problem, we perform a set of analytical evaluations and compare the results with a traditional finite element (FE) solution. The simulations will be performed at steady state conditions following a harmonic excitation and the comparison will be made in terms of the predicted displacement amplitude $u(x)$ as a function of the spatial location. 

Before proceeding with the analysis of the numerical results, we recall that the fractional order $\alpha$ was found to be a complex quantity. A complex-order derivative of a real-valued function is a complex-valued function \cite{Valerio,Hartley}, therefore $u(x)$ will also be complex. Hollkamp \cite{Hollkamp} discussed this aspect and concluded that, for harmonic excitations, the time response of a complex fractional oscillator is an analytic function, therefore justifying the physical response being represented by $u(t)=\textrm{Re}[\bar{u}(t)$]. Thus, in the following plots, the curves corresponding to the analytical solutions are the real parts of the calculated complex-valued function.

In the following, four different forcing frequencies will be considered (namely 100, 500, 1000, 3100 Hz). The selection of these values was dictated by the specific position of the pass bands and band gaps for the selected periodic rod. More specifically, $\Omega=100$ Hz corresponds to a low frequency regime where the classical homogenization ``rule of mixtures" is a valid approach (the wavelength to unit cell size ratio is $\lambda/a \approx 20$). The frequency $\Omega=500$ Hz is still located in the first pass band, however it approaches the limit of the homogenization assumption ($\lambda/a \approx 4$). The frequency $\Omega=1000$ Hz is of particular interest since it is located in the first band gap where the wave is attenuated and the homogenization assumption starts breaking down \cite{Wu}. The frequency $\Omega=3100$ Hz is located in a higher frequency band gap (the third for our system) and denotes a regime were the wave is attenuated and the homogenization assumption is certainly not applicable. 

Each of the following benchmark studies will compare traditional FE results with the analytical solutions obtained using the exponential and the Mittag-Leffler kernels. Both approaches for the value of $\bar{c}$ will also be considered. In the following plots, results obtained using the constant value of $\bar{c}$ as described in \S \ref{sec:conC} will be labeled as ``$\bar{c}$ approach'' while those obtained using the frequency-dependent speed given in \S \ref{sec:varC} will be labeled as ``$\bar{c}_\omega$ approach." The FE results were obtained by performing steady state analyses using a commercial package (COMSOL Multiphysics). The specific values of $\bar{c}$ and $\alpha$ for each of the selected frequencies are obtained from phase one and summarized in Tables \ref{tab1}. 

\begin{table}[!htbp] 
\begin{center}
\caption{Values of $\bar{c}$ and $\alpha$ following the approaches given in \S \ref{sec:conC} \& \S \ref{sec:varC}.}\label{wall}
\begin{tabular}{|c|c|c|c|c|c|}
\hline $\Omega$ (Hz) & $\bar{c}$ & $\alpha$ using $\bar{c}$ & $\bar{c}_\omega$ & $\alpha$ using $\bar{c}_\omega$ \\
\hline 100&3980.4&2.0003 + 0.0003i&3978.2&2\\
\hline 500&3980.4&2.0031 + 0.0226i&3909.2&2\\
\hline 1000&3980.4&1.7583 - 0.0408i&4000&1.7569 - 0.0357i\\
\hline 3100&3980.4&1.9895 - 0.0550i&4133.3&1.9662 - 0.0307i\\
\hline
\end{tabular} \label{tab1}
\end{center}
\end{table}

Concerning the FE numerical simulations, a harmonic displacement of amplitude $u_o$=1 mm was applied at $x$=0. A low reflecting boundary condition was applied at the right end ($x$ = 400 m) to model an infinite structure. 

\subsection{Exponential solution} \label{sec:infexpsol}

The exponential solution for the semi-infinite rod is given by Eq.~\eqref{eq:entiresolM}. Fig.~\ref{fig:expinf} presents the spatial solution from Eq.~\eqref{eq:entiresolM} for the four selected values of $\Omega$ and compares them with the reference FE solution. In all plots, the results using the ``$\bar{c}$ approach" and the ``$\bar{c}_\omega$ approach" are exactly equivalent. Thus, the exponential solution does not show any sensitivity to the methodology used to select $\bar{c}$. This is also confirmed by the value of the stability parameters $a\pi-b\textrm{ln}(\lambda)$ (see Eq. \eqref{eq:stabcond}) which returns exactly the same value for both approaches. 

The analytical solutions for $\Omega$=100 Hz match the finite element results quite well with an RMS error of 0.018\% over the domain, as seen in  Fig.~\ref{fig:expinf}a. At 500 Hz (Fig.~\ref{fig:expinf}b), the analytical solutions match the phase of the finite element results, but do not match the amplitudes locally. This type of behavior is not unexpected because the frequency is near the long wavelength limit so that the response is increasingly dominated by scattering effects. Like other homogenization techniques, the proposed fractional homogenization model cannot capture localized effects due to wave scattering and interference; however, it allows modeling the wave behavior within the band gap largely beyond the homogenization limit. 

Consider the plot at $\Omega$ = 1000 Hz (Fig.~\ref{fig:expinf}c) located in the first band gap. The analytical solution agrees very well with the finite element results with an RMS error as low as 0.0005\%. The amplitude of the wave decays in space. Note that classical homogenization techniques cannot capture at all the behavior within a band gap and cannot provide an analytical solution. This is a remarkable advantage of the proposed technique. Lastly, consider the plots at $\Omega$ = 3100 Hz in Fig.~\ref{fig:expinf}d. The analytical solution in this third band gap matches well the plot from FE simulations, except for a few locations (some have a percent difference error as high as 50\% between the analytical and numerical solutions). This discrepancy shows, once again, that the fractional homogenization technique is quite effective in the frequency band gaps. 

As Makris noted, the imaginary part of a fractional derivative affects the phase modulation of a function \cite{Makris}. Considering the ``$\bar{c}$ approach", if we were to set Im($\alpha$) in Table \ref{tab1} equal to zero and plot the solutions using only the real part of the calculated order $\alpha$, we would see a distinct phase mismatch between the analytical and numerical solutions. This supports the fact that the imaginary part of a fractional derivative corresponds to controlling the phase of a function. In addition, the imaginary part can also have an effect on the stability. Considering Eq.~\eqref{eq:stabcond}, the value of $b$ (related to the imaginary part of $\alpha$) plays a significant role in determining whether or not Eq.~\eqref{eq:stabcond} is satisfied. While Makris is correct in stating that the physical meaning of the imaginary part of a fractional order corresponds to phase modulation, it appears that this is not the only significance of the imaginary part. Together, the values of the real and imaginary parts of a fractional order determine the stability.  

\begin{figure}[h!] 
  \begin{center}
    \centerline{\includegraphics[scale=0.65,angle=0]{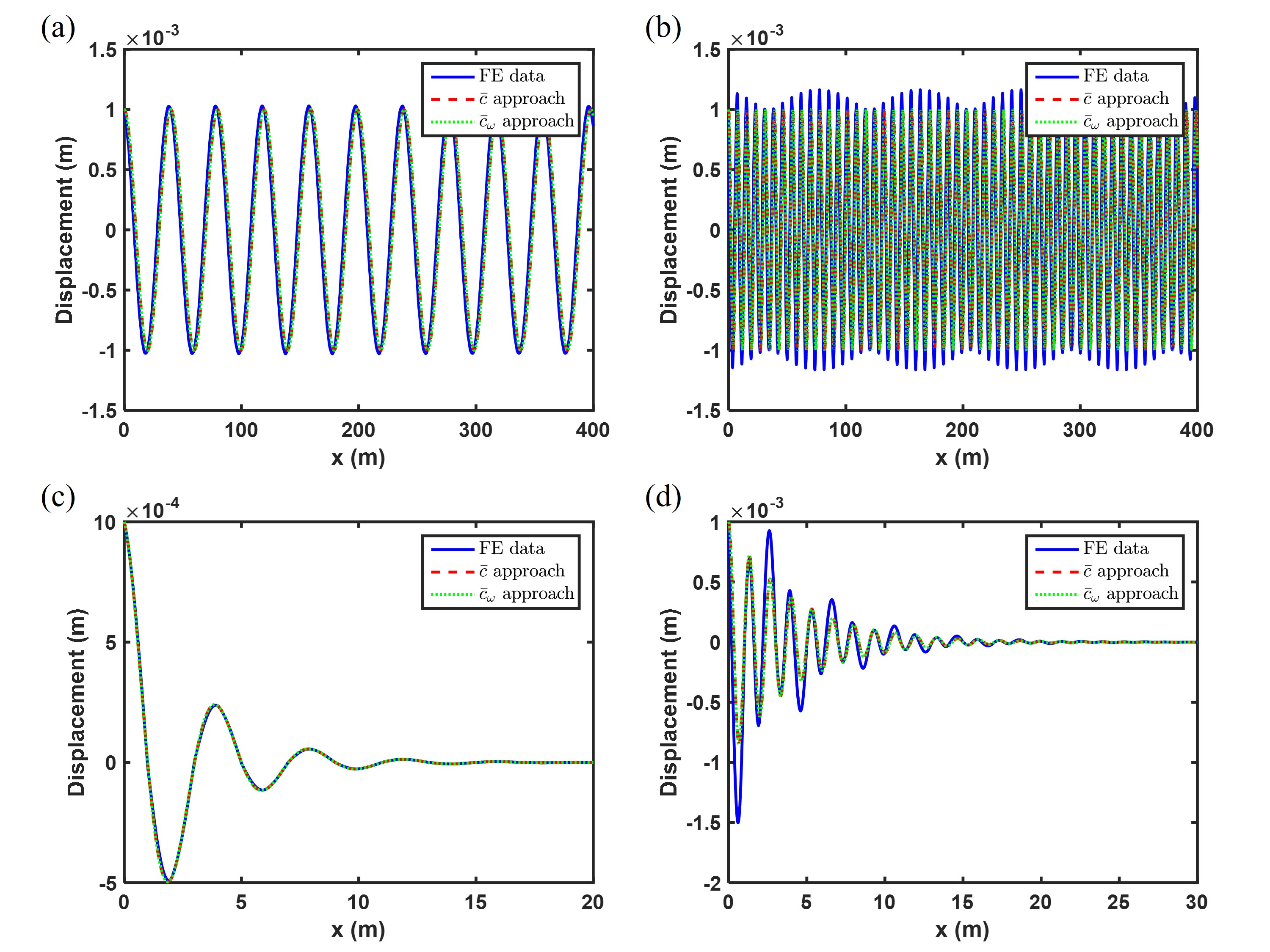}}
    \caption{The solution to the fractional wave equation when using the exponential solution form for a forcing frequency of a) $\Omega$ = 100 Hz, b) $\Omega$ = 500 Hz, c) $\Omega$ = 1000 Hz, d) $\Omega$ = 3100 Hz. }
    \label{fig:expinf}
  \end{center}
\end{figure}

\subsection{Mittag-Leffler solution} \label{sec:infMLsol}

The Mittag-Leffler solution for the semi-infinite rod is given by Eq.~\eqref{eq:solABC} where $B$ is given by Eq.~\eqref{eq:BMI}. As previously mentioned, the Mittag-Leffler functions in Eq. \eqref{eq:solABC} are unstable whenever Re$(\alpha)>2$ \cite{Diethelm}, no matter the value of the imaginary part. As a result, the solutions using the constant $\bar{c}$ might be unstable at any frequency where Re$(\alpha)>2$. However, we see from Fig.~\ref{fig:MLinf} that the results using $\bar{c}$ are in fact stable when Re$(\alpha)>2$ (except in Fig.~\ref{fig:MLinf}d). In actuality, both of the terms in Eq.~\eqref{eq:solABC} are growing without bound; however, the value of $B$ actually causes the instabilities of the two terms to cancel each other out. Furthermore, the first three plots in Fig.~\ref{fig:MLinf} show that the results using the two approaches for $\bar{c}$ are equivalent. However, if the value of $B$ were determined in a different manner than in Eq.~\eqref{eq:BMI} or if the value of $B$ in Eq.~\eqref{eq:BMI} is rounded off prematurely, then we would observe an instability arising in the constant $\bar{c}$ approach. This is exactly what is seen in Fig.~\ref{fig:MLinf}d. The instability was verified to be caused by the imaginary part of $\alpha$ rather than the real part. As suggested earlier, the imaginary part of a fractional order does play a role in the stability. Unfortunately, an expression of the stability (like Eq.~\eqref{eq:stabcond} for the exponential solutions) could not be obtained for Mittag-Leffler functions with complex orders. More mathematical theory needs to be studied and developed for complex order Mittag-Leffler functions before a stability expression can be obtained. 

Extreme caution must be used whenever it is observed that Re$(\alpha)>2$ when using Mittag-Leffler solution kernels. It is recommended that the ``$\bar{c}_\omega$ approach" be implemented when using Mittag-Leffler functions to avoid a potential instability in the analytical solution since the fractional order never possesses a value where Re$(\alpha)>2$. This is precisely why we considered the frequency-dependent fractional wave speed approach.

Comparing the exponential and Mittag-Leffler solutions show that both sets are nearly identical. In fact, the plots in Figs.~\ref{fig:MLinf}a and \ref{fig:MLinf}b are exactly equivalent to Figs.~\ref{fig:expinf}a and \ref{fig:expinf}b. Comparing Figs.~\ref{fig:MLinf}c and ~\ref{fig:expinf}c shows that the solution using the exponential function was more accurate at $\Omega=1000$ Hz than the Mittag-Leffler solution. The evaluation of the Mittag-Leffler function was subject to some intrinsic error that the exponential solution did not possess (see \S \ref{sec:error}). Although the exponential and Mittag-Leffler solutions in Figs.~\ref{fig:expinf} and \ref{fig:MLinf} are nearly alike, this is not always the case. Either the exponential or the Mittag-Leffler solution will be better suited depending on whether the bar is infinite or finite and depending on specific boundary conditions.  

\begin{figure}[h!] 
  \begin{center}
    \centerline{\includegraphics[scale=0.65,angle=0]{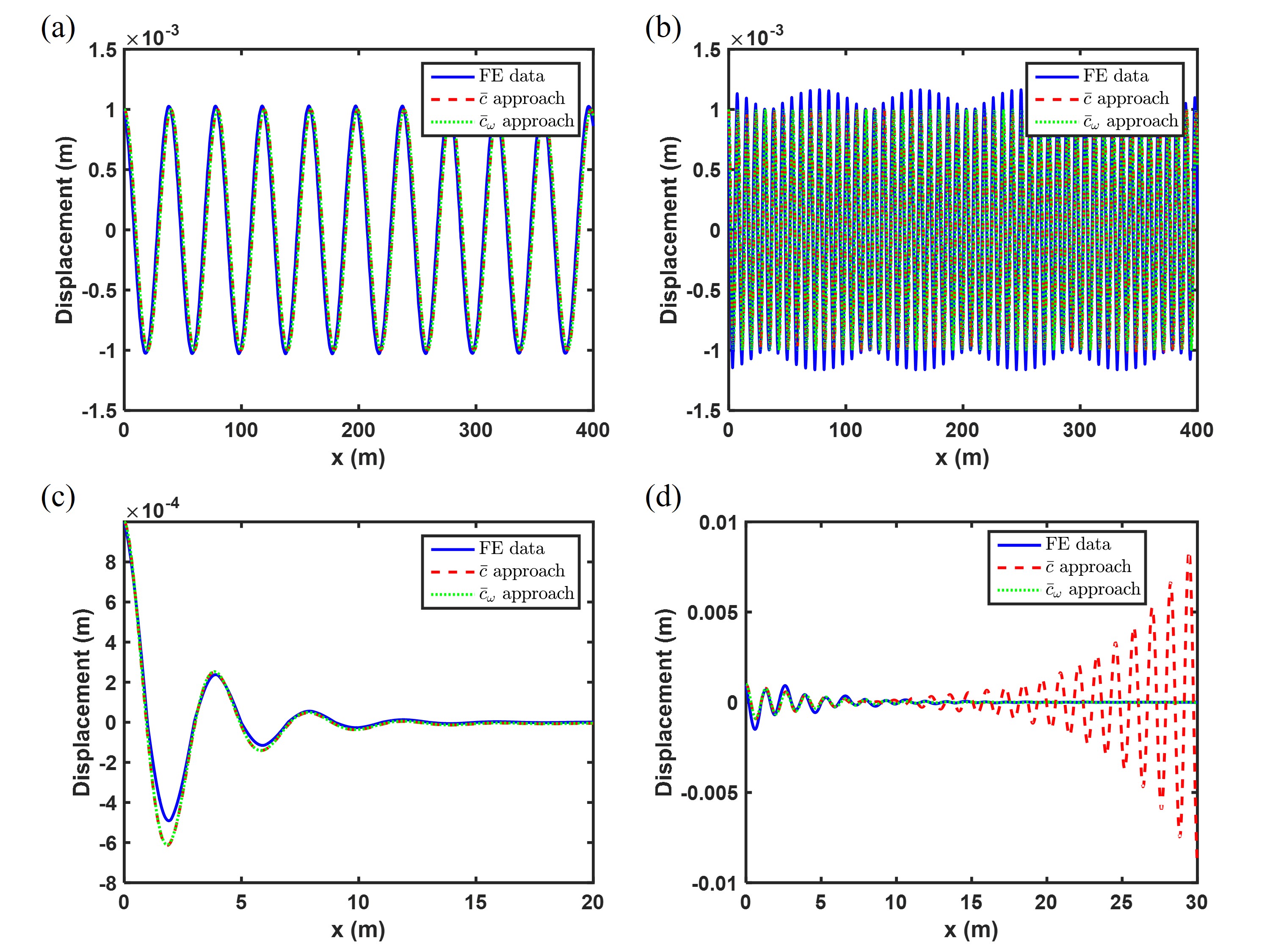}}
    \caption{The solution to the fractional wave equation when using the Mittag-Leffler solution form for a forcing frequency of a) $\Omega$ = 100 Hz, b) $\Omega$ = 500 Hz, c) $\Omega$ = 1000 Hz, d) $\Omega$ = 3100 Hz.}
    \label{fig:MLinf}
  \end{center}
\end{figure}

\subsection{Sources of Error} \label{sec:error}

It was already mentioned that the proposed technique is effectively a homogenization approach and, as such, it is locally accurate only for long wavelengths. When approaching short wavelength regimes, local interference effects cannot be accurately captured. Nevertheless, this approach is able to correctly capture the wave attenuation as well as the proper spatial phase. In addition, it provides a fully analytical solution to predict the wave behavior. 

Another error source is due to the evaluation of the Mittag-Leffler function itself. As pointed out by Garrappa \cite{Garrappa}, while the Mittag-Leffler function plays a fundamental role in fractional calculus, there are surprisingly very few methods available for its numerical evaluation. The evaluation of the infinite summation term in Eq.~\eqref{eq:Mittag} is not a trivial task and is actually an ongoing area of computational research \cite{Garrappa,Garrappa2}. A few common MATLAB functions do exist \cite{CodeGarrappa,CodePodlubny} to numerically evaluate the Mittag-Leffler function, but they are not valid for complex values of $\alpha$ and $\beta$. In order to evaluate the Mittag-Leffler function for complex orders, we developed a modified version of \cite{CodeGarrappa}. After numerical investigations, it was determined that our edited version of Garrappa's code had some intrinsic errors associated with it. In fact, this error associated with the evaluation of the complex-order Mittag-Leffler function contributes to some of the slight amplitude overshoots and phase mismatches such as those seen in Fig.~\ref{fig:MLinf}c.

\section{Conclusions} \label{sec:conclusion}

This paper presented a methodology to develop a space-fractional wave equation capable of capturing the dynamic response of one-dimensional periodic structures. This methodology could be effectively interpreted as a homogenization technique. From a mathematical perspective, this approach converts a wave equation with spatially-variable coefficients (i.e. the classical differential model for a periodic medium) into a space-fractional differential equation with constant coefficients. The order $\alpha$ of the fractional operator was obtained by imposing that the dispersion behavior of both the fractional and the integer-order wave equations were equivalent. The technique yielded a complex and frequency dependent fractional order $\alpha$. It was found that the real part of $\alpha$ is related to the decay or growth of the solution while the imaginary part provides a frequency-dependence of the phase and influences the stability.

Overall, the proposed fractional approach allowed 1) the development of a homogenization method capable of capturing the response within frequency bandgaps (that is beyond the classical long wavelength limit), and 2) deriving a closed-form, analytical solution of the approximate dynamic response of the inhomogeneous periodic medium. The fact that the fractional models led to an analytical solution that was a good approximation of the periodic, bi-material rod showcased the remarkable potential of the use of fractional operators for continuum mechanics applications. The ability of this method to provide closed-form solutions could also have important implications for inverse problems in material design or even sensing. Furthermore, while this work addressed one-dimensional systems, it is conceivable that this technique could be extended to higher dimensions. 

The importance of the ability of fractional homogenized models to capture the system response within the band gaps cannot be overstated. This is a strength of the proposed modeling approach over classical homogenization techniques that rely strictly on integer order operators and cannot capture the decaying nature of the wave propagation in these regimes without adding artificial damping terms. Although the proposed approach clearly has high-frequency capabilities, it is not necessarily free of error at all frequencies and cannot capture local effects due to high frequencies interference. Nonetheless, as more complicated structures possessing multiple frequency band gaps continue to be explored in design and research, this fractional homogenization technique will be a key tool to studying the dynamic response of these structures.

\section{Acknowledgments}
This work was partially supported by the Air Force Office of Scientific Research under YIP \# FA9550-15-1-0133 and by the National Science Foundation under DCSD CAREER $\#1621909$.

\appendix
\section{Basic Definitions of Fractional Calculus}\label{App_A}

Several different definitions of a fractional derivative exist. Perhaps the two most common definitions are the Riemann-Liouville and the Caputo derivatives. Contrary to integer-order derivatives, fractional derivatives are nonlocal in nature. Just like integration, the value of a fractional derivative is dependent on its bounds, or terminals. A fractional derivative calculated from a lower bound is a ``left-handed'' derivative while a fractional derivative calculated from an upper bound is a ``right-handed'' derivative.

The left-handed Riemann-Liouville fractional derivative of order $\alpha$ and lower bound $a$ is

\begin{eqnarray}
 \leftidx{_a^{RL}}{D}{_t^\alpha} f(t) = \frac{1}{\Gamma(n-\alpha)}\frac{\textrm{d}^n}{\textrm{d}t^n}\int_{a}^{t} f(\tau)(t-\tau)^{n-\alpha -1} \textrm{d}\tau
\end{eqnarray}

\noindent where $\Gamma(t)$ is the Gamma function and $n$=$\ceil{\alpha}$. The right-handed Riemann-Liouville fractional derivative of order $\alpha$ and upper bound $b$ is

\begin{eqnarray}
\leftidx{_t^{RL}}{D}{_b^\alpha} f(t) = \frac{(-1)^n}{\Gamma(n-\alpha)}\frac{\textrm{d}^n}{\textrm{d}t^n}\int_{t}^{b} f(\tau)(t-\tau)^{n-\alpha -1} \textrm{d}\tau.
\end{eqnarray}

\noindent The left-handed Caputo fractional derivative is

\begin{eqnarray}
\leftidx{_a^C}{D}{_t^\alpha}  f(t) = \frac{1}{\Gamma(n-\alpha)}\int_{a}^{t} \frac{\textrm{d}^n f(\tau)}{\textrm{d}\tau^n} (t-\tau)^{n-\alpha -1} \textrm{d}\tau.
\end{eqnarray}

\noindent The right-handed Caputo fractional derivative is

\begin{eqnarray}
\leftidx{_t^C}{D}{_b^\alpha}  f(t) = \frac{(-1)^n}{\Gamma(n-\alpha)}\int_{t}^{b} \frac{\textrm{d}^n f(\tau)}{\textrm{d}\tau^n} (t-\tau)^{n-\alpha -1} \textrm{d}\tau.
\end{eqnarray}

\noindent We now explore the Laplace transforms of the left-handed derivatives with a lower terminal of 0. The Laplace transform of Riemann-Liouville fractional derivative is

\begin{eqnarray} \label{LTRL}
\int_{0}^{\infty} \textrm{e}^{-st} \Big(\leftidx{_0^{RL}}{D}{_t^\alpha}  f(t)\Big) \textrm{d}t = s^{\alpha}F(s) - \sum_{k=0}^{n-1} s^k \Big[\leftidx{_0^{RL}}{D}{_t^{\alpha-k-1}}  f(t)\Big]_{t=0}, \qquad n-1 < \alpha \leq n
\end{eqnarray}

\noindent while the Laplace transform of Caputo fractional derivative is

\begin{eqnarray} \label{LTC}
\int_{0}^{\infty} \textrm{e}^{-st} \Big(\leftidx{_0^{C}}{D}{_t^\alpha}  f(t)\Big) \textrm{d}t = s^{\alpha}F(s) - \sum_{k=0}^{n-1} s^{\alpha -k-1} f^{(k)}(0), \qquad n-1 < \alpha \leq n
\end{eqnarray}

\noindent where $f^{(k)}(0)$ is the $k^{th}$ order derivative of $f$ evaluated at $t=0$. Note, as reflected by Eq.~\eqref{LTC}, the Caputo derivative uses the same initial values that a typical integer order problem does (first derivative, second derivative, etc). Meanwhile, the initial values of the Riemann-Liouville definition are actually non-integer order derivative values of the function at $t=0$. The physical meaning of the necessary initial conditions using the Riemann-Liouville definition is an open question. On the other hand, the Caputo derivative lends itself to initial values which have a well-defined physical interpretation (initial position, velocity, acceleration, etc).

\section{Mittag-Leffler solution when $\alpha=2$}\label{App_B}

Here, we look to show that the Mittag-Leffler solution is equivalent to the exponential solution when $\alpha$=2. To do so, first note a property relating the Mittag-Leffler function to the cosine and sine functions. Let $\theta$ be a generic constant. Mittag-Leffler properties related to cosine and sine include

\begin{equation} \label{eq:MLcosine}
E_{2,1}(-\theta^2 x^2) = \textrm{cos}(\theta x),
\end{equation}

\begin{equation} \label{eq:MLsine}
\theta x E_{2,2}(-\theta^2 x^2) = \textrm{sin}(\theta x).
\end{equation}

Let us now start with the spatial part of the exponential solution (Eq.~\eqref{eq:Meerspace}) and substitute $\alpha$ = 2 to obtain

\begin{equation} 
\phi(x) = u_o\textrm{e}^{i\frac{\omega}{\bar{c}}x}
\end{equation}

or, in terms of sine and cosine,

\begin{equation} \label{eq:Meeral2}
\phi(x) = u_o\textrm{cos}(\frac{\omega}{\bar{c}}x)+\textrm{i}u_o\textrm{sin}(\frac{\omega}{\bar{c}}x).
\end{equation}

Now, consider the spatial part of the Mittag-Leffler solution (Eq.~\eqref{eq:spacesol}) with $\alpha$=2. It is

\begin{eqnarray} \label{eq:spatialML}
\phi(x) = A E_{2,1}(-\frac{\omega^2}{\bar{c}^2} x^2)+B x E_{2,2}(-\frac{\omega^2}{\bar{c}^2} x^2).
\end{eqnarray}

The value of $A$ is $u_o$. The value of $B$ is given by Eq.~\eqref{eq:BMI} with $c$=$\bar{c}$. Using Eqs.~\eqref{eq:MLcosine} and \eqref{eq:MLsine}, this can be written as

\begin{equation} \label{eq:Cal2}
\phi(x) = u_o\textrm{cos}(\frac{\omega}{\bar{c}}x)+\textrm{i}u_o \textrm{sin}(\frac{\omega}{\bar{c}}x)
\end{equation}

\noindent which is equivalent to Eq.~\eqref{eq:Meeral2}.


\bibliographystyle{unsrt} 
\bibliography{report} 
\end{document}